\documentclass[twocolumn]{autart}    

\usepackage{natbib}
\usepackage{color}
\usepackage{graphicx}         
\usepackage{amsmath} 
\usepackage{amssymb}  
\usepackage{url}
\usepackage{subfigure}
\usepackage{amstext, enumerate}

\newtheorem{assmpt}{\bf Assumption}
\newtheorem{rem1}{\bf Remark}

\newcommand{\R}{\ensuremath{{\mathbb R}}}

\newcommand{\cG}{\ensuremath{{\mathcal G}}}
\newcommand{\EE}{\ensuremath{{\mathcal E}}}

\newcommand{\NN}{\ensuremath{{\mathcal N}}}
\newcommand{\XX}{\ensuremath{{\mathcal X}}}
\newcommand{\LLL}{\ensuremath{{\mathcal L}}}

\newcommand{\cC}{\ensuremath{{\mathcal C}}}

\newcommand{\ddd}{\ensuremath{{\mathsf d}}}

\newcommand{\LL}{\ensuremath{{\mathbb L}}}

\begin{document}

\begin{frontmatter}
\title{Initialization-free Privacy-guaranteed Distributed Algorithm for Economic Dispatch Problem\thanksref{footnoteinfo}}

\thanks[footnoteinfo]{This research was supported by Korea Electric Power Corporation through Korea Electrical Engineering \& Science Research Institute (grant number: R15XA03-47) and by the National Research Foundation of Korea (NRF) grant funded by the Korea government (Ministry of Science and ICT) (No. NRF-2017R1E1A1A03070342).}
		
\author[SS]{Hyeonjun Yun}\ead{hjyun.w@gmail.com},    
\author[SNU]{Hyungbo Shim}\ead{hshim@snu.ac.kr},
\author[GIST]{Hyo-Sung Ahn}\ead{hyosung@gist.ac.kr} 
		
\address[SS]{Mechatronics R\&D Center, Samsung Electronics Co., Ltd., 1-1 Samsungjeonja-ro, Hwaseong-si, Gyeonggi-do 18448, Korea}
\address[SNU]{ASRI, Department of Electrical and Computer Engineering, Seoul National University, Seoul, Korea.}  
\address[GIST]{School of Mechanical Engineering, Gwangju Institute of Science and Technology, Gwangju, Korea.}

\begin{keyword}                           
Economic dispatch, Power grids, Distributed optimization, Multi-agent systems, Synchronization.
\end{keyword}                             

\begin{abstract}
This paper considers the economic dispatch problem for a network of power generators and customers.
In particular, our aim is to minimize the total generation cost under the power supply-demand balance and the individual generation capacity constraints.
This problem is solved in a distributed manner, i.e., a dual gradient-based continuous-time distributed algorithm is proposed in which only a single dual variable is communicated with the neighbors and no private information of the node is disclosed. 
The proposed algorithm is simple and no specific initialization is necessary, and this in turn allows on-line change of network structure, demand, generation constraints, and even the participating nodes.
The algorithm also exhibits a special behavior when the problem becomes infeasible so that each node can detect over-demand or under-demand situation of the power network.
Simulation results on IEEE 118 bus system confirm robustness against variations in power grids. 
\end{abstract}
		
\end{frontmatter}

\section{Introduction}\label{sec:introduction}
	
The smart grid will become more decentralized with the integration of distributed energy resources (DER), storage devices, and customers. 
The three important key features of the smart grid are large scale of components, highly variable nature of DER, and dynamic network topology. 
In view of optimization, these three features make the traditional centralized optimization techniques impractical, and pose a need to develop distributed methods in grid optimization problems. 
These observations lead us to design distributed solutions for the economic dispatch problem (EDP), where a group of power generators attempts to achieve power supply-demand balance while minimizing the total generation cost (i.e., sum of the individual costs) and complying with individual generation capacity constraints. 
	
Early solutions for the EDP have been developed in a centralized manner such as lambda-iteration \citep{Z09}, Lagrangian relaxation \citep{GHO96}, genetic algorithm \citep{BPK94}, and so on. 
Then, a lot of research effort has been devoted to obtain distributed algorithmic solutions for the EDP due to the distributed nature of the future smart grid.
In particular, discrete-time consensus-based algorithms have been the majority of the distributed strategies for the EDP reported in the literature. 
Many works have considered convex quadratic objective functions for the power generation cost \citep{YTX13,KHMM14,EE15}, but most of them require an initialization process because of the usage of decaying stepsizes \citep{KHMM14,YLWWSMJ17}, sequential algorithmic steps \citep{XMFL15}, or a reset rule \citep{YTX13}. 
Meanwhile, some recent works have proposed continuous-time consensus-based solutions, which allow to use the classical stability analysis for the convergence of the proposed algorithms. 
\cite{AKLO17} have considered optimal power generation and distribution, but need an initialization process and do not consider capacity constraints. 
\cite{CC15} proposed algorithms which use Laplacian-nonsmooth-gradient dynamics with dynamic average consensus, and the requirement of initialization is overcome by \cite{CC16}. 
\cite{YHL16} also presented initialization-free algorithms which combine the concept of projected gradient dynamics with dynamic average consensus.
	
The purpose of this paper is to propose a new continuous-time distributed algorithm to solve the EDP. 
The features of the proposed algorithm can be stated as follows.
When we formulate the dual problem from the primal one (as in \citep{Simon16}), only the single equality constraint is contained in the Lagrangian while the other constraints are considered when the distributed dual function is constructed. 
As a result, solving the dual problem becomes as simple as integrating just a first-order differential equation.
It will be seen that the first-order differential equation is decentralized in the sense that it is a simple sum of different vector fields. 
Then, inspired by the recent result on practical synchronization based on an average of vector fields, studied by \cite{KYSKS16}, we develop a fully distributed continuous-time algorithm, which does not require exchange of private information such as power generation cost, generation capacity, and power demand.	
Moreover, the distributed algorithm does not need any initialization process, which allows on-line changes of DER, loads, network topology and so on. 
The cost to pay for these benefits is that the optimal solution is obtained approximately but not exactly. 
However, we will show that the power supply-demand balance is always satisfied even if the solution is approximately optimal, and that the approximation error can be made arbitrarily small to satisfy desired precision by taking sufficiently large coupling gain.
Another advantage of the proposed algorithm is that, when the EDP is infeasible, the solution of the proposed algorithm shows a special divergent behavior, from which each node can detect infeasibility and figure out the amount of shortage or surplus of the total demand.
This may be an interesting observation since handling infeasible situation is rare in the literature to the authors' knowledge.

The proposed algorithm is presented in Section \ref{sec:main} based on some basics reviewed in Section \ref{sec:cent}.
The algorithm is tested in Section \ref{sec:example} for IEEE 118 bus system, and more discussions about the proposed method will follow in Section \ref{sec:remarks}.

\noindent{\em Notation:} 
We denote $1_N = [1,\ldots,1]^T\in \R^N$. 
For a vector $x$ and a matrix $A$, $|x|$ and $|A|$ denote the Euclidean norm and its induced matrix norm, respectively. 
A function $f:\R \rightarrow \R$ is monotonically increasing (decreasing) if $f(x)\leq f(y)$ ($f(x)\geq f(y)$) for all $x$ and $y$ such that $x \leq y$. 
Now we present basic notions and results from algebraic graph theory \citep{BCM09}.
An undirected graph used in this paper is a pair $\cG = (\NN,\EE)$, where $\NN = \{1,2,\ldots,N\}$ is a node set and $\EE \subseteq \NN \times \NN$ is an edge set such that $(i,j) \in \EE$ if and only if $(j,i) \in \EE$.
The (symmetric) Laplacian matrix $\LL=[l_{ij}]\in \R^{N \times N}$ is defined as $l_{ij}=-1$ if there is an edge between the nodes $i$ and $j$ ($i \not = j$), $l_{ij}=0$ otherwise, and $l_{ii} := -\sum_{j \not = i} l_{ij}$.
If $\cG$ is connected, then $0$ is a simple eigenvalue of $\LL$

\section{Problem formulation}\label{sec:p_f}

The economic dispatch problem (EDP) of interest in this paper is formulated as a convex optimization problem with equality and inequality constraints as follows:
\begin{subequations}\label{eq:EDP}
\begin{align}
\min_{x_1,x_2,\cdots,x_N} &\quad \sum_{i=1}^N J_i(x_i) \label{eq:EDP_cost} \\	\text{subject to} &\quad \sum_{i=1}^{N}  x_i = \sum_{i=1}^{N}  d_i, \label{eq:EDP_eq}\\
&\quad \underline{x}_i \leq x_i \leq \bar{x}_i, \quad \forall i =1,\ldots,N. \label{eq:EDP_ineq}
\end{align}
\end{subequations}
In the above, it is supposed that there are $N$ nodes, and each node has its own power generation $x_i$, local objective function $J_i:\R \rightarrow \R$ representing the cost of power generation, $d_i$ is the local power demand, and $\bar x_i$ and $\underline x_i$ are the upper and the lower limits of node $i$'s power generation, respectively.
The task is to minimize the total cost \eqref{eq:EDP_cost} by determining $x_1, \cdots, x_N$ under two constraints; the supply-demand balance equation \eqref{eq:EDP_eq} and the generation capacity inequality \eqref{eq:EDP_ineq} (for details, refer to, e.g., \cite{WW12} and \cite{KHMM14}).
In particular, we assume that the information of $J_i(\cdot)$, $d_i$, $\bar x_i$, and $\underline x_i$ are private for the node $i$ so that each node does not want to disclose them to other nodes.
	
\begin{assmpt}\label{assmpt:cost}
The local objective function $J_i:\R \rightarrow \R$ is $\cC^2$ (twice continuously differentiable) and strictly convex for all $i=1,\ldots,N$.
\end{assmpt}

When node $i$ has no generator (e.g., a customer node that has demand only), we take  $\underline{x}_i=\bar{x}_i=0$, and choose an arbitrary local objective function $J_i(\cdot)$ satisfying Assumption \ref{assmpt:cost} such that $J_i(0)=0$.
For the node that has no load, simply take $d_i = 0$ with suitable $J_i(\cdot)$, $\bar x_i$, and $\underline x_i$.
We say that the EDP \eqref{eq:EDP} is {\em feasible} if $\sum_{i=1}^{N} \underline{x}_i  \leq \sum_{i=1}^N d_i \leq \sum_{i=1}^{N} \bar{x}_i$, otherwise it is {\em infeasible}. 
We also say that the EDP is of over-demand if $\sum_{i=1}^{N} \bar{x}_i < \sum_{i=1}^N d_i$ and is of under-demand if $\sum_{i=1}^N d_i < \sum_{i=1}^{N} \underline{x}_i$.

\section{Preliminary: a centralized solution}\label{sec:cent}
	
In this section, we review the standard (centralized) procedure to solve the EDP with emphasis on a few key ingredients that will be used in the distributed solution of the next section.
	
Let $x := [x_1,x_2,\cdots,x_N]^T$, $\XX_i := \{ x_i \in \R : \underline x_i \le x_i \le \bar x_i \}$, and $\XX := \XX_1 \times \cdots \times \XX_N$.
From the optimization problem \eqref{eq:EDP}, we consider the following Lagrangian
\begin{equation}\label{eq:Lag_ineq}
\LLL(x,\lambda) = \sum_{i=1}^N J_i(x_i) + \lambda \left(\sum_{i=1}^{N} d_i - \sum_{i=1}^{N} x_i \right) 
\end{equation}	
where $\lambda \in \R$ is the Lagrange multiplier associated with \eqref{eq:EDP_eq}.
Then, the Lagrange dual function $g:\R \rightarrow\R$ is obtained as
\begin{align*}
\begin{split}
g(\lambda) &= \inf_{x \in \XX} \mathcal{L}(x, \lambda) = \inf_{x \in \XX} \sum_{i=1}^N \left(J_i(x_i) + \lambda(d_i-x_i) \right) \\
&= \sum_{i=1}^N \inf_{x_i \in \XX_i} \left(J_i(x_i) + \lambda(d_i-x_i) \right) =: \sum_{i=1}^N g_i(\lambda)
\end{split}
\end{align*}
where the third equality holds thanks to the distributed nature of the problem.
Let us call $g_i:\R \rightarrow \R$ a {\em distributed} dual function. 
Note that the inequality constraints \eqref{eq:EDP_ineq} are not included in the Lagrangian \eqref{eq:Lag_ineq} and instead the dual function $g$ (and thus, the distributed dual functions $g_i$ as well) is obtained in consideration of the constraints \eqref{eq:EDP_ineq}.

The analytic form of $g_i(\lambda) = \inf_{\underline x_i \le x_i \le \bar x_i} J_i(x_i) + \lambda(d_i-x_i)$ can be obtained as follows.
First, note that, for a given $\lambda$, the derivative of the cost with respect to $x_i$, $(dJ_i/dx_i)(x_i) - \lambda$, is a strictly increasing function of $x_i$ by Assumption \ref{assmpt:cost}.
Hence, if $(dJ_i/dx_i)(\bar x_i) - \lambda < 0$, then the cost is decreasing on the interval $[\underline x_i, \bar x_i]$ and thus, achieves its minimum at $x_i = \bar x_i$.
Similarly, if $(dJ_i/dx_i)(\underline x_i) - \lambda > 0$, then the cost is increasing on $[\underline x_i, \bar x_i]$ and the minimum occurs at $x_i = \underline x_i$.
Finally, if $(dJ_i/dx_i)(\underline x_i) \le \lambda \le (dJ_i/dx_i)(\bar x_i)$, then the minimum is achieved on $[\underline x_i, \bar x_i]$ where it holds that
\begin{equation}\label{eq:vi}
\frac{dJ_i}{dx_i}(x_i) = \lambda.
\end{equation}
Let the solution of \eqref{eq:vi} be $v_i(\lambda)$ where $v_i(\cdot)$ is the inverse function of $(dJ_i/dx_i)(\cdot)$ which is well-defined, $\cC^1$, and strictly increasing because of Assumption \ref{assmpt:cost}.
Therefore, the explicit form of the distributed dual function $g_i$ is given by
\begin{equation}
g_i(\lambda) = J_i(\theta_i(\lambda)) + \lambda(d_i - \theta_i(\lambda))
\end{equation}
where $\theta_i:\R \rightarrow \R$ is defined as
\begin{equation}\label{eq:gicases}
\theta_i(\lambda) = \begin{cases}
\underline x_i, & \lambda < \frac{dJ_i}{dx_i}(\underline x_i), \\
v_i(\lambda), & \frac{dJ_i}{dx_i}(\underline x_i) \le \lambda \le \frac{dJ_i}{dx_i}(\bar x_i), \\
\bar x_i, & \frac{dJ_i}{dx_i}(\bar x_i) < \lambda.
\end{cases}
\end{equation}
Note that these $\theta_i(\lambda)$'s minimize $\LLL$ for the given $\lambda$.
Moreover, the concave and $\cC^1$ property of the dual function $g_i(\lambda)$ follows from \citep[Prop~6.1.1]{Bert}.

\begin{rem1}\label{rem:nogen1}
When $\underline{x}_i = \bar{x}_i$ for the node~$i$ (i.e., the node~$i$ produces fixed amount of power or has no generator if $\underline x_i=\bar x_i=0$), it is seen that $g_i(\lambda) = \inf_{x_i \in \XX_i} J_i(x_i) + \lambda(d_i-x_i) = J_i(\bar{x}_i) + \lambda(d_i-\bar{x}_i)$ for all $\lambda \in \R$, which confirms \eqref{eq:gicases} as well.
\end{rem1}

With the dual function $g$, the Lagrange dual problem \citep{BV04} of the EDP \eqref{eq:EDP} is obtained as a form of unconstrained optimization problem:
\begin{equation}\label{eq:dual_problem}
\max_\lambda \quad g(\lambda)=\sum_{i=1}^N g_i(\lambda).
\end{equation}
Here, it is noted that each $g_i$ has its derivative as 
\begin{equation}\label{eq:CT_GD_cent_dg_i}
\frac{dg_i}{d \lambda}(\lambda) = d_i - \theta_i(\lambda)
\end{equation}
which is {\em continuous, monotonically decreasing, and uniformly bounded.}
In case that $\underline x_i = \bar x_i$, we have $(dg_i/d\lambda)(\lambda) = d_i - \bar x_i$.
Hereafter, we investigate the property of the dual problem \eqref{eq:dual_problem}.
Let us define
\begin{equation}\label{eq:lambda1}
\underline \lambda := \min_{i\in \NN} \frac{dJ_i}{dx_i}(\underline x_i) \le \max_{i \in \NN} \frac{dJ_i}{dx_i}(\bar x_i) =: \bar \lambda.
\end{equation}
Then, the function $(dg/d\lambda)(\lambda)$, which is also continuous and monotonically decreasing, satisfies 
\begin{equation}\label{eq:dg_upperbound}
\frac{d g}{d \lambda}(\lambda) = \begin{cases}
\frac{d g}{d \lambda}(\underline{\lambda}) = \sum_{i=1}^N d_i - \sum_{i=1}^N \underline x_i, & \forall \lambda \leq \underline \lambda, \\
\frac{d g}{d \lambda}(\bar{\lambda}) = \sum_{i=1}^N d_i - \sum_{i=1}^N \bar x_i, & \forall \lambda \ge \bar{\lambda} .
\end{cases}
\end{equation}
Suppose that the EDP \eqref{eq:EDP} is feasible.
Then, $(dg/d\lambda)(\lambda) \ge 0$, $\forall \lambda \le \underline{\lambda}$, and $(dg/d\lambda)(\lambda) \le 0$, $\forall \bar{\lambda} \le \lambda$.
Hence, there exists a nonempty connected closed interval $\Lambda^* \subset \R$ (that is possibly unbounded or a point) such that $(dg/d\lambda)(\lambda) = 0$, $\forall \lambda \in \Lambda^*$.
Therefore, the Lagrangian dual problem \eqref{eq:dual_problem} achieves its maximum at all points $\lambda^* \in \Lambda^*$.
Moreover, the optimal solution $x_i^*$ of the primal problem \eqref{eq:EDP} is obtained through \eqref{eq:gicases} from any dual optimal solution $\lambda^* \in \Lambda^*$ of the dual problem \eqref{eq:dual_problem} as
\begin{equation}\label{eq:opt_gen_ineq}
x_i^* = \theta_i(\lambda^*), \quad \forall i \in \NN.
\end{equation}
This is because each cost function $J_i$ is convex, the equality constraint \eqref{eq:EDP_eq} is affine, and the set $\XX_1 \times \cdots \times \XX_N$ is polyhedral, so that there is no duality gap between the primal problem \eqref{eq:EDP} and the dual problem \eqref{eq:dual_problem} when the EDP \eqref{eq:EDP} is feasible \citep{BNO03}. 
Note that $x_i^*$ is uniquely defined for all $\lambda^* \in \Lambda^*$ because, if $\Lambda^*$ is not a single point so that $\lambda^*$ is not unique, it means that $\lambda^* \not \in \{ \lambda : \exists i \text{ s.t. } (dJ_i/dx_i)(\underline x_i) < \lambda < (dJ_i/dx_i)(\bar x_i) \}$ by the construction of \eqref{eq:CT_GD_cent_dg_i} through \eqref{eq:gicases}, and therefore, again by \eqref{eq:gicases}, $\theta_i(\lambda^*)$ has the same value on $\Lambda^*$.

One way to compute an optimal solution $\lambda^* \in \Lambda^*$ of \eqref{eq:dual_problem} is to use a classical approach of the gradient descent algorithm.
For this, let us denote by $\lambda(t)$ the (time-varying) estimate of $\lambda^*$ which obeys
\begin{equation}\label{eq:CT_GD_cent}
\dot \lambda(t) = \frac{d g}{d \lambda}(\lambda(t)) = \sum_{i=1}^N \frac{d g_i}{d \lambda}(\lambda(t)) .
\end{equation}
If the EDP is feasible, it is obvious that $\lambda(t)$ converges to the set $\Lambda^*$ as time tends to infinity from any initial condition $\lambda(0) \in \R$, because $(dg/d\lambda)(\lambda)>0$ if $\lambda$ is less than the minimum (if exists) of the interval $\Lambda^*$, and $(dg/d\lambda)(\lambda)<0$ if $\lambda$ is greater than the maximum (if exists) of $\Lambda^*$.
If the EDP is not feasible, for example, if it is of over-demand, then by \eqref{eq:dg_upperbound} and by the fact that $(dg/d\lambda)(\cdot)$ is monotonically decreasing, we have that $(dg/d\lambda)(\lambda) \ge (dg/d\lambda)(\bar\lambda) = \sum_{i=1}^N d_i - \sum_{i=1}^N \bar x_i > 0$, $\forall \lambda \in \R$.
This means that $\lambda(t)$ diverges to $+\infty$.
On the other hand, if the EDP is of under-demand, one can similarly show that $\lambda(t)$ diverges to $-\infty$.

\section{A distributed solution}\label{sec:main}
	
In this section, we present a distributed solution for the EDP \eqref{eq:EDP}. 
The idea is inspired by the observation that $(dg/d\lambda)$ in \eqref{eq:CT_GD_cent} is decomposed as a sum of $(dg_i/d\lambda)$, and by the recent result of \citep{KYSKS16} that can estimate a solution to the average of different vector fields in a multi-agent system.
The proposed solution is that each node $i \in \NN$ runs the following dynamics
\begin{equation}\label{eq:consensus_rule}
\dot \lambda_i(t) = \frac{d g_i}{d \lambda}(\lambda_i(t)) + k \sum_{j \in \NN_i} (\lambda_j(t) - \lambda_i(t))
\end{equation}
with a common coupling gain $k>0$.
Here, $\lambda_i(t) \in \R$ is the internal state of the individual node $i$ and $\NN_i$ is the index set of neighboring nodes of the node $i$.
It will be shown that we can make $\lambda_i(t)$ converge to arbitrarily small neighborhood of $\Lambda^*$ (so that $\theta_i(\lambda_i(t))$ will become a sufficiently rich approximate of the optimal solution $x_i^*$) under the following assumption:
\begin{assmpt}\label{assmpt:graph}
The graph $\cG$ is undirected and connected.
\end{assmpt}
In the assumption, the graph $\cG$ implies the communication graph over the power network, which may be different from the power transmission lines.
Since it is a communication network, it is not unrealistic to assume it is `undirected.'

It is noted from \eqref{eq:consensus_rule} that there is no centralized server and each node just communicates their own $\lambda_i$ with its neighboring nodes.
No private information such as $J_i$, $\underline x_i$, $\bar x_i$, and $d_i$ are exchanged, and the function $\theta_i$ as well as the function $g_i$ (both of which are computed from $J_i$, $\underline x_i$, $\bar x_i$, and $d_i$) are kept within the node $i$.

It will turn out that the {\em distributed solution} $\theta_i(\lambda_i(t))$ is a sub-optimal solution because it approximates $x_i^*$ but may not be the same.
However, even in this case, the following theorem shows that the supply-demand balance \eqref{eq:EDP_eq} is satisfied, which is of utmost important in practice.

\begin{thm}\label{thm:cs_arb_k} 
Suppose that the EDP \eqref{eq:EDP} under Assumptions \ref{assmpt:cost} and \ref{assmpt:graph} is feasible. 
Then, for any $k>0$ and any $\lambda_i(0) \in \R$, the solution $\lambda_i(t)$ of \eqref{eq:consensus_rule} satisfies $\lim_{t \rightarrow \infty} \dot \lambda_i(t) = 0$ for all $i \in \NN$, and 
\begin{equation}\label{eq:equal}
\lim_{t \rightarrow \infty} \sum_{i=1}^N \theta_i({\lambda}_i(t)) = \sum_{i=1}^N d_i.
\end{equation}
\end{thm}

Before presenting a proof of Theorem \ref{thm:cs_arb_k}, let us develop a representation of \eqref{eq:consensus_rule} in another coordinates, which all the forthcoming analyses are based on.
Let $\boldsymbol{\lambda} := [\lambda_1,\ldots,\lambda_N]^T$ and $\boldsymbol{f}(\boldsymbol{\lambda}) := [(dg_1/d\lambda)(\lambda_1),\ldots,(dg_N/d\lambda)(\lambda_N)]^T$.
Then, the system \eqref{eq:consensus_rule} can be written simply as 
\begin{equation}\label{eq:limiting_eq}
\dot{\boldsymbol{\lambda}} = \boldsymbol{f}(\boldsymbol{\lambda}) - k \LL \boldsymbol{\lambda} =: \boldsymbol{F}(\boldsymbol{\lambda})
\end{equation}
where $\LL$ is the Laplacian matrix representing the graph $\cG$.
Choose any orthonormal matrix $U \in \R^{N \times N}$ whose first row is $(1/\sqrt{N})1_N^T$.
Let $W := (1/\sqrt{N})U$, then $$W = \begin{bmatrix} \frac1N 1_N^T \\ R^T \end{bmatrix}, \qquad
W^{-1} = \sqrt{N}U^T = \begin{bmatrix} 1_N & Q \end{bmatrix}$$
where $R \in \R^{N \times (N-1)}$ and $Q \in \R^{N \times (N-1)}$.
By construction, we have that $Q = NR$, $Q^TQ = NI_{N-1}$, and $R^TQ = I_{N-1}$.
Now, by the coordinate transformation 
\begin{equation}\label{eq:CT}
\begin{bmatrix} \xi_1 \\ \xi_e \end{bmatrix} = W \boldsymbol{\lambda} 
= \begin{bmatrix} \frac{1}{N} 1_N^T \boldsymbol{\lambda} \\ R^T \boldsymbol{\lambda} \end{bmatrix}
\end{equation}
where $\xi_e \in \R^{N-1}$, it is seen that $\boldsymbol{\lambda} = W^{-1} [\xi_1, \xi_e^T]^T$, or, $\lambda_i = \xi_1 +  Q_i \xi_e$ where $Q_i$ is the $i$-th row of $Q$.
Moreover, the system \eqref{eq:consensus_rule} is transformed into
\begin{subequations}\label{eq:CT_xi}
\begin{align}
\dot{\xi}_1 &= \frac1N 1_N^T \boldsymbol{f}(1_N\xi_1 + Q\xi_e) = \frac{1}{N} \sum_{i=1}^N \frac{d g_i}{d \lambda} \left( \xi_1 + Q_i \xi_e \right) \notag \\
&= \frac1N \frac{dg}{d\lambda}(\xi_1) + \frac1N \tilde{f}(\xi_e; \xi_1), \label{eq:slow} \\
\dot{\xi_e} &= -kR^T \LL Q \xi_e + R^T \boldsymbol{f} \left(1_N \xi_1  + Q \xi_e\right) \label{eq:fast}
\end{align}
where
\begin{equation*}
\tilde{f}(\xi_e; \xi_1)	:= \sum_{i=1}^N \left( \frac{d g_i}{d \lambda} (\xi_1+ Q_i \xi_e) - \frac{d g_i}{d \lambda}(\xi_1 ) \right). 
\end{equation*}
\end{subequations}
It should be noted that the matrix $R^T \LL Q$ is symmetric and all its eigenvalues are positive real numbers, whose smallest one is denoted by $\sigma_2$. 
Moreover, from the definitions of $R$ and $Q$, it can be shown that $\sigma_2$ is actually the smallest non-zero eigenvalue of $\LL$ under Assumption \ref{assmpt:graph} \citep[Theorem 1.37]{BCM09}.
Note also that the vector field $\boldsymbol{f}$ is uniformly bounded, and thus, define $b_{\boldsymbol{f}} := \max_{\boldsymbol{\lambda}\in\R^N} |\boldsymbol{f}(\boldsymbol{\lambda})|$.

\begin{pf}
First of all, we claim that the solution of \eqref{eq:CT_xi} is {\em bounded} for any initial condition and for any $k > 0$.
Boundedness of $\xi_e(t)$ of \eqref{eq:fast} follows from the facts that $-R^T \LL Q$ is Hurwitz and that $\boldsymbol{f}$ is uniformly bounded.
It can be also seen that $\xi_1(t)$ of \eqref{eq:slow} cannot become unbounded because, by \eqref{eq:CT_GD_cent_dg_i}, \eqref{eq:gicases}, and the feasibility assumption, we have, for sufficiently large $\xi_1$, $\dot \xi_1 = (1/N) \sum_{i=1}^N (d_i - \bar x_i)$ is non-positive and, for sufficiently small $\xi_1$, $\dot \xi_1 = (1/N) \sum_{i=1}^N (d_i - \underline x_i)$ is non-negative.

Now, with boundedness of $\xi_1(t)$ and $\xi_e(t)$, we apply LaSalle's invariance principle \citep{K02} for their convergence.
Define a $\cC^1$ function $Y(\boldsymbol{\lambda}) = -\sum_{i=1}^N g_i(\lambda_i) + \frac12 k \boldsymbol{\lambda}^T \LL \boldsymbol{\lambda}$. 
Then, from \eqref{eq:limiting_eq}, its time derivative becomes
\begin{align*}
\dot{Y}(\boldsymbol{\lambda}) = - \boldsymbol{F}^T(\boldsymbol{\lambda}) \cdot \dot{\boldsymbol{\lambda}}  = - |\boldsymbol{F}(\boldsymbol{\lambda})|^2 \leq 0.
\end{align*}
Therefore, LaSalle's invariance principle asserts that $\xi_1(t)$ and $\xi_e(t)$ converge to (the largest invariance set in) the set $E := \{ \boldsymbol{\lambda} : \boldsymbol{F}(\boldsymbol{\lambda}) = 0 \}$.
Since $\dot{\boldsymbol{\lambda}} = 0$ on the set $E$, we have $\lim_{t \rightarrow \infty} \dot{\boldsymbol{\lambda}}(t) = \lim_{t \rightarrow \infty} \boldsymbol{F}(\boldsymbol{\lambda}(t)) = 0$.
Moreover, since $\lim_{t \rightarrow \infty} 1_N^T \boldsymbol{F}(\boldsymbol{\lambda}(t)) = 0$, it follows that
\begin{align*}
&\lim_{t \rightarrow \infty} 1_N^T \boldsymbol{F}(\boldsymbol{\lambda}(t)) = \lim_{t \rightarrow \infty} \left( 1_N^T \boldsymbol{f}({\boldsymbol{\lambda}}(t)) - k 1_N^T \LL {\boldsymbol{\lambda}}(t) \right) \\
&= \lim_{t \rightarrow \infty}\sum_{i=1}^N \frac{dg_i}{d\lambda}(\lambda_i(t)) = \lim_{t \rightarrow \infty} \sum_{i=1}^N \left(d_i - \theta_i(\lambda_i(t))\right) = 0
\end{align*}
which concludes the proof.
\hfill $\Box$
\end{pf}
	
The following theorem asserts that the optimal solution $x_i^*$ can be approximated by $\theta_i(\lambda_i(t))$ with arbitrarily small error within a finite time when $k$ is large.

\begin{thm}\label{thm:cs}
Suppose that the EDP \eqref{eq:EDP} under Assumptions \ref{assmpt:cost} and \ref{assmpt:graph} is feasible. 
Then, for any $\epsilon>0$, there exists $\bar{k}>0$ such that for all $k\ge\bar{k}$, each solution $\lambda_i(t)$ of \eqref{eq:consensus_rule}, with $\lambda_i(0) \in \R$, $\forall i \in \NN$, satisfies $\limsup_{t\to\infty}|\theta_i(\lambda_i(t)) - x_i^*| \le \epsilon$.
In particular, if the initial conditions satisfy $\underline \lambda \le \lambda_i(0) \le \bar{\lambda}$, $\forall i \in \NN$, then there is a non-increasing function $T(\cdot)$ such that
\begin{equation}\label{eq:goestooptimal}
\big| \theta_i (\lambda_i(t)) - x_i^* \big| \leq \epsilon, \qquad \forall t \ge T(k).
\end{equation}
\end{thm}

\begin{pf}
Let us first suppose that $\underline \lambda \le \lambda_i(0) \le \bar{\lambda}$, $\forall i \in \NN$.
It is noted that the continuous function $(dg_i/d\lambda)(\lambda)$ in \eqref{eq:CT_GD_cent_dg_i} is uniformly continuous because $\theta_i(\cdot)$ is constant except on the compact interval $\left[(dJ_i/dx_i)(\underline x_i), (dJ_i/dx_i)(\bar x_i)\right]$ where $\theta_i$ is continuous. 
Therefore, one can choose $\delta>0$ such that, $\forall a, b \in \R$,
\begin{equation}\label{eq:delta}
|a - b| \le \delta \quad \Rightarrow \quad 
\left| \frac{dg_i}{d\lambda}(a) - \frac{dg_i}{d\lambda}(b) \right| \le \frac{\epsilon}{3N}, \; \forall i \in \NN.
\end{equation}
Define
\begin{equation}\label{eq:bark}
\bar k := \frac{2b_{\boldsymbol{f}}}{\sigma_2\delta} . 
\end{equation} 
Let $V_e(\xi_e) = (1/2) |\xi_e|^2$.
Then, it follows from \eqref{eq:fast} and $|R| = 1/\sqrt{N}$ that, for $k \ge \bar k$,
\begin{align*}
\dot V_e &\le -k \xi_e^T R^T \LL Q \xi_e + \xi_e^T R^T \boldsymbol{f}(1_N\xi_1 + Q \xi_e) \\
&\le -k\sigma_2 |\xi_e|^2 + \frac{b_{\boldsymbol{f}}}{\sqrt{N}} |\xi_e| \\
&\le - \frac{k\sigma_2}{2} |\xi_e|^2 - \frac{\bar k \sigma_2}{2} |\xi_e| \left( |\xi_e| - \frac{2b_{\boldsymbol{f}}}{\bar k \sigma_2 \sqrt{N}} \right).
\end{align*}
This implies that $\dot V_e \le -k\sigma_2 V_e$ if $|\xi_e| \ge \delta/\sqrt{N}$; that is, $|\xi_e(t)| \le \exp(-(k\sigma_2/2) t) |\xi_e(0)|$ as long as $|\xi_e(t)| \ge \delta/\sqrt{N}$.
Since $|\xi_e(0)| \le |R^T| |\boldsymbol{\lambda}(0)| \le (1/\sqrt{N})\sqrt{N} \max\{|\underline \lambda|,|\bar{\lambda}|\} =: M_\lambda$, we have
\begin{equation}\label{eq:T1}
|\xi_e(t)| \le \frac{\delta}{\sqrt{N}}, \quad \forall t \ge T_1(k) := \frac{2}{k\sigma_2} \ln \frac{\sqrt{N} M_\lambda}{\delta}
\end{equation}
(take $T_1(k)=0$ if $\sqrt{N}M_\lambda < \delta$).
Moreover, for $t \ge T_1(k)$, we have that $|Q_i\xi_e(t)| \le |Q_i| |\xi_e(t)| \le \sqrt{N} (\delta/\sqrt{N}) = \delta$.
With \eqref{eq:delta}, we then have
\begin{multline}\label{eq:3e}
|\tilde f(\xi_e;\xi_1)| = \left| \sum_{i=1}^N \left( \frac{dg_i}{d\lambda}(\xi_1+Q_i\xi_e) - \frac{dg_i}{d\lambda}(\xi_1) \right) \right| \\
\le N \frac{\epsilon}{3N} = \frac{\epsilon}{3} .
\end{multline}
Now, define
$$\Lambda_{\epsilon}^* := \left\{ \lambda \in \R: \left|\frac{dg}{d\lambda}(\lambda)\right| \leq \frac{2\epsilon}{3} \right\}$$
which includes the set $\Lambda^*$.
We will show that there exists $T_2(k) \ge 0$ such that the solution $\xi_1(t)$ of \eqref{eq:slow} belongs to the set $\Lambda_{\epsilon}^*$ for $t \ge T_1(k)+T_2(k)$. 
For this, we claim that, after the time $T_1(k)$, the state $\xi_1(t)$, if located outside of the set $\Lambda_{\epsilon}^*$, approaches $\Lambda_{\epsilon}^*$ with the speed at least $\epsilon/(3N)$.
Indeed, since $(dg/d\lambda)(\lambda)$ is monotonically decreasing, $(dg/d\lambda)(\lambda) < -2\epsilon/3$ outside of $\Lambda_{\epsilon}^*$ to the right in $\R$, and $(dg/d\lambda)(\lambda) > 2\epsilon/3$ outside of $\Lambda_{\epsilon}^*$ to the left (while there may be the cases where no outside of $\Lambda_{\epsilon}^*$ to the left/right exists if $\Lambda_{\epsilon}^*$ is unbounded).
With \eqref{eq:slow} and \eqref{eq:3e}, this justifies the claim. 
On the other hand, under the feasibility condition, it follows from \eqref{eq:gicases}, \eqref{eq:CT_GD_cent_dg_i}, and \eqref{eq:slow} that $|\dot \xi_1(t)| = |(1/N) \sum_{i=1}^N (d_i - \theta_i(\xi_1(t) + Q_i \xi_e(t)))| \le (1/N) \sum_{i=1}^N (\bar x_i - \underline x_i)$ for any $t$.
Thus, even if $\xi_1(0) = (1/N)\sum_{i=1}^N \lambda_i(0) \in [\underline \lambda, \bar{\lambda}]$, the state $\xi_1(T_1(k))$ may be located outside of $[\underline \lambda, \bar{\lambda}]$ up to the distance of $(T_1(k)/N) \sum_{i=1}^N (\bar x_i - \underline x_i)$.
Then, since $\Lambda_{\epsilon}^* \cap [\underline \lambda,\bar{\lambda}]$ is not empty by the feasibility (see \eqref{eq:dg_upperbound}), the state $\xi_1(t)$, started from $\xi_1(T_1(k))$, arrives at the set $\Lambda_{\epsilon}^*$ within the time
$$T_2(k) := \frac{3N}{\epsilon} \left( \bar \lambda - \underline \lambda + \frac{T_1(k)}{N} \sum_{i=1}^N (\bar x_i - \underline x_i) \right).$$
Since $\lambda_i = \xi_1 + Q_i \xi_e$, for all $t \ge T_1(k)+T_2(k) =: T(k)$,
\begin{multline*}
\left|\frac{dg}{d\lambda}(\lambda_i(t))\right| \le \left| \frac{dg}{d\lambda}(\xi_1(t)) \right| \\
+ \left| \frac{dg}{d\lambda}(\xi_1(t)+Q_i\xi_e(t)) - \frac{dg}{d\lambda}(\xi_1(t)) \right| \le \frac{2\epsilon}{3} + \frac{\epsilon}{3} \le \epsilon
\end{multline*}
for all $i \in \NN$ by similar reasoning to \eqref{eq:3e}.
Therefore,
\begin{multline*}
|\theta_i(\lambda_i(t))- \theta_i(\lambda^*)| \leq \sum_{j =1}^N |\theta_j (\lambda_i(t)) - \theta_j(\lambda^*) | \\
= \left| \sum_{j=1}^N (\theta_j(\lambda_i(t)) - \theta_j(\lambda^*)) \right| = \left| -\frac{dg}{d\lambda}(\lambda_i(t)) \right| \le \epsilon
\end{multline*}
where the first equality holds by the fact that $\theta_j(\cdot)$ is monotonically increasing for all $j \in \NN$, and the second equality holds from \eqref{eq:CT_GD_cent_dg_i}, \eqref{eq:opt_gen_ineq}, and the supply-demand balance.

For the case that $\lambda_i(0) \in \R$, $\forall i \in \NN$, the proof is similarly done taking into account that the value of $M_\lambda$ in \eqref{eq:T1} can be arbitrarily large.
\hfill $\Box$
\end{pf}

On top of Theorem \ref{thm:cs}, the following corollary specifies the behavior of the proposed algorithm \eqref{eq:consensus_rule} in the case that the EDP \eqref{eq:EDP} is infeasible.

\begin{cor}\label{thm:cs_inf}
Suppose that the EDP \eqref{eq:EDP} under Assumptions \ref{assmpt:cost} and \ref{assmpt:graph} is infeasible. 
Then, there exists $T^\dagger>0$ such that each solution $\lambda_i(t)$ of \eqref{eq:consensus_rule} with $k\ge\bar k$, initiated as $\underline \lambda \le \lambda_i(0) \le \bar{\lambda}$, $\forall i \in \NN$, satisfies 
\begin{align}
&\begin{cases} \lambda_i(t) > \bar \lambda, & \text{(over-demand)} \\ \lambda_i(t) < \underline \lambda, & \text{(under-demand)} \end{cases} \quad \text{for $t > T^\dagger$, and}  \label{eq:aaa} \\
&\lim_{t \to \infty} \dot \lambda_i(t) = \begin{cases} \frac1N \sum_{i=1}^N (d_i - \bar x_i), & \text{(over-demand)} \\ \frac1N \sum_{i=1}^N (d_i - \underline x_i), &\text{(under-demand)}. \end{cases} \label{eq:bbb}
\end{align}
In fact, $\lambda_i(t) \to \infty$ (over-demand), or $\lambda_i(t) \to -\infty$ (under-demand) as $t\to\infty$.
If $\lambda_i(0) \in \R$, $\forall i \in \NN$, then the same holds but the time $T^\dagger$ can be arbitrarily large.
\end{cor}

\begin{pf}
(This proof continues the proof of Theorem \ref{thm:cs}.)
From \eqref{eq:gicases}, \eqref{eq:CT_GD_cent_dg_i} and \eqref{eq:slow}, it is obvious that $\dot \xi_1(t) \le (1/N) \sum_{i=1}^N (d_i - \underline x_i) =: M_{\sf u} < 0$ in the case of under-demand, or $\dot \xi_1(t) \ge (1/N) \sum_{i=1}^N (d_i - \bar x_i) =: M_{\sf o} > 0$ in the case of over-demand.
Since $\xi_1(0) \in [\underline \lambda,\bar{\lambda}]$, after the time $T^\dagger := \max\{(\bar{\lambda}-\underline{\lambda}+\delta)/\min\{M_{\sf o},-M_{\sf u}\}, T_1(\bar k)\}$, the state $\xi_1(t)$ for $t > T^\dagger$ is either less than $\underline \lambda-\delta$ or greater than $\bar{\lambda}+\delta$.
Since $\lambda_i(t)=\xi_1(t) + Q_i \xi_e(t)$ and $|Q_i \xi_e(t)| \le \delta$ for $t > T^\dagger$, $\forall i \in \NN$, the statement \eqref{eq:aaa} follows.
Also, the last statement follows since $\xi_1(t) \to \pm \infty$ as $t\to\infty$, depending on the cases.
Finally, if $\lambda_i(t) \not \in [\underline \lambda,\bar{\lambda}]$, then $\theta_i(\lambda_i(t))$ is either $\bar x_i$ or $\underline x_i$ depending on the cases.
Then, it is seen from \eqref{eq:fast} that $\xi_e(t)$ converges to an equilibrium because $\boldsymbol{f}(1_N\xi_1(t)+Q\xi_e(t))$ becomes a constant vector for $t \ge T^\dagger$, so that $\lim_{t \to \infty} \dot \xi_e(t) = 0$.
Therefore, the statement \eqref{eq:bbb} follows from \eqref{eq:slow} because $\lim_{t \to \infty} \dot \lambda_i(t) = \lim_{t \to \infty} (\dot \xi_1(t) + Q_i \dot \xi_e(t))$.
For the case that $\lambda_i(0)\in\R$, $\forall i \in \NN$, the proof is the same except that $T_1(\bar k)$ is arbitrarily large, so that the claim follows.
\hfill $\Box$
\end{pf}

A message from Theorem \ref{thm:cs} and Corollary \ref{thm:cs_inf} is that the selection of the gain $k$ in \eqref{eq:consensus_rule} has much freedom as long as it is sufficiently large.
However, too large $k$ is not desirable since it makes the algorithm sensitive to communication noise and makes the discretization finer when \eqref{eq:consensus_rule} is implemented in a digital computer.
We illustrate a way to choose $k$ (and $\bar k$ as well) in Section \ref{sec:remarks}.

\begin{rem1}
The dynamics \eqref{eq:consensus_rule} corresponds to the update rule presented by \cite{NO09} if \eqref{eq:consensus_rule} is discretized.
Indeed, by forward difference method with the sampling period $\tau$, the dynamic equation \eqref{eq:consensus_rule} becomes
\begin{equation*}
\lambda_i^{\ddd} (n+1) =  \lambda_i^{\ddd} (n) + \tau k \sum_{j \in \NN_i}  (\lambda_j^{\ddd}(n)-\lambda_i^{\ddd}(n)) + \tau \frac{d g_i}{d \lambda}(\lambda_i^{\ddd}(n))
\end{equation*}
where $\lambda_i^{\ddd}(n)=\lambda_i(n\tau)$, which corresponds to the form of \cite[eq.~(3)]{NO09}.
It is clearly seen that we use a constant stepsize while there are many results that use decaying stepsizes in the literature.
Since the algorithms with decaying stepsizes exhibit different behavior in response to on-line changes in the network as time goes on, they are not initialization-free algorithms (see Section \ref{sec:remarks}). 
Finally, we recall that, by resorting to continuous-time dynamics \eqref{eq:consensus_rule}, it was possible to employ well-known classical stability results such as LaSalle's invariance principle in the proof of Theorem \ref{thm:cs_arb_k}.
\end{rem1}

\section{Discussions}\label{sec:remarks}

In order to be applied in real applications, a distributed algorithm to solve EDP should have a few desirable properties as follows.

{\em Decentralized design and initialization-free operation:} 
In practice, a power network is time-varying one in the sense that the demand $d_i$, or the individual generation cost $J_i$ can be changed from time to time depending on the owner's decision of node $i$. 
The number of node $N$ can also be changed if a new node joins the network or a node leaves it. 
Since these changes are not able to be detected by all nodes in the network at a time, it is not desirable to ask each node to do something in response to the local change in the network.
Instead, it is better for the algorithm to run continuously without any special treatment even if such changes occur (which we call {\em initialization-free} property).
In addition, when a new node is joining the network for example, it is desirable that the design of the algorithm in the new node does not need much global information (i.e., information about the network topology and/or all other nodes), which we call {\em decentralized design}.
The proposed algorithm \eqref{eq:consensus_rule} achieves both properties to some extent because the only global information is the gain $k$ (which encapsulates all other global information).
We illustrate an idea of computing $k$ at the end of this section, from which it is supposed that the network operator announces the value of $k$ in public {\em a priori}.
Then, a newcomer to the network just computes two functions $g_i$ and $\theta_i$ from its own local information (by \eqref{eq:gicases} and \eqref{eq:CT_GD_cent_dg_i}) and joins the network with its own dynamics \eqref{eq:consensus_rule}.
Also, when any changes occur in the node~$i$ during operation, the node can simply re-computes $g_i$ and $\theta_i$ and continues its operation seamlessly.

{\em Privacy-guaranteed:} 
Since the information such as $J_i$, $d_i$, $\bar x_i$, and $\underline x_i$ may be private, it is not desirable to send them to other nodes or the center. 
The proposed algorithm \eqref{eq:consensus_rule} exchanges only the single variable $\lambda_i$ and keeps the privacy.

{\em Time for trustful solution:}
Since the optimization is solved by iteration, it is desirable to know {\em a priori} how long it takes to obtain a reasonable solution after the transient caused by an on-line change.
Theorem \ref{thm:cs} and Corollary \ref{thm:cs_inf} suggest the worst case of required time as $T(k)$ and $T^\dagger$, as long as the initial conditions $\lambda_i(0)$ belongs to the finite interval $[\underline \lambda, \bar{\lambda}]$ of \eqref{eq:lambda1}. 
(To enjoy this property, the network operator needs to announce in public the values of $\underline \lambda$ and $\bar \lambda$ as well so that a newcomer can set its initial condition accordingly.)
From the proof of Theorem \ref{thm:cs}, it is clear that $\lambda_i(t)$ belongs to this interval in most time of normal operation except for the short time period of $T_1(k)$ after a change.
So, if the changes of the network are not too frequent, the proposed times are valid.
However, the suggested values of $T(k)$ and $T^\dagger$ are conservative, and it is the future work to find tighter upper bounds of them.
On the other hand, since the change can happen at any time and it is not easy for each node to detect the abrupt change, a question arises: when can one trust the value of $\lambda_i(t)$?
An idea is to synchronize the changes in time over the network; for example, any changes in the cost, demand, or the network can only occur at the multiple of $T_{\sf sync}$, where $T_{\sf sync} > \max \{T(k), T^\dagger\}$.
Then, one can use the value $\lambda_i(t)$ at every multiple of $T_{\sf sync}$ to determine the power generation.

{\em Fast convergence:}
It is desired to obtain a reasonable solution quickly.
Related to this, it is noted that the time $t$ in the algorithm \eqref{eq:consensus_rule} is not the real world time but the computer time in the communication network.
Therefore, with sufficiently fast computers and communications, the times $T(k)$ and $T^\dagger$ can be reached faster in real world time.
Moreover, if one introduces a scaling factor $\alpha>1$ to \eqref{eq:consensus_rule} like $(d\lambda_i/dt) = \alpha (dg_i/d\lambda)(\lambda_i) + \alpha k \sum_{j\in\NN_i}(\lambda_j-\lambda_i)$, then the operation is accelerated (without changing the proofs in this paper; that is, by dividing both sides by $\alpha$, the time index $t$ now becomes $(\alpha t)$).

{\em Exact solution:} 
It is desirable for a distributed algorithm to find the solution of \eqref{eq:EDP} exactly, but the proposed algorithm just approximates it (although arbitrary small error of the solution is achieved in finite time by increasing $k$).
To solve this problem, the approaches of, e.g., \cite{SLWY15,QL17,HCIL17} may be helpful, which is our future work.

{\em Supply-demand balance:} 
The balance constraint \eqref{eq:EDP_eq} is a key constraint in power networks, and the proposed algorithm always guarantees it (by Theorem \ref{thm:cs_arb_k}) even if the solution is approximate one.
While some error in the equality of \eqref{eq:equal} of Theorem \ref{thm:cs_arb_k} is inevitable in finite time, the maximum of the error is simply given by $\epsilon N$ at the finite time $T(k)$ of Theorem \ref{thm:cs}, due to \eqref{eq:goestooptimal}, and the error goes to zero as time tends to infinity.

{\em Simplicity:} 
Computational burden in each node should not be high.
Compared from the algorithms by \cite{CC16} and \cite{YHL16} which require three-dimensional dynamics for each node, the proposed algorithm only requires one-dimensional dynamics \eqref{eq:consensus_rule} and a static map \eqref{eq:opt_gen_ineq}, which is relatively simple to be implemented. 

{\em Communication delay:}
In practice, communication incurs a delay.
If $\lambda_j$ of other nodes are delayed in \eqref{eq:consensus_rule}, then too large $k$ may lead to instability in general.
Robustifying the algorithm against the delay is beyond the scope of this paper, and is left as a future work.

{\em Choosing a suitable $k$:}
In order to implement the proposed algorithm \eqref{eq:consensus_rule} at each node, suitable $k$ needs to be chosen.
The problem is that, while it is enough to choose $k \ge \bar k$, the value of $\bar k$ depends on global information as seen in \eqref{eq:bark} and the global information may vary as time goes on. 
One heuristic solution is to choose $\bar k$ by repeated simulations for various scenarios, based on the reasoning that sufficiently large $k$ always does the job.
Another way is to pick the worst case value of $\bar k$ under the assumptions that (a) the network has maximum capacity (that is, there is an upper bound $N_{\max}$ of the number $N$ of participating nodes), (b) the cost function $J_i$ belongs to a finite collection (that is, there is a pre-determined set of candidate functions $\{\hat J_1, \hat J_2, \cdots, \hat J_n\}$ and each node simply chooses one of them), (c) there are lower/upper bounds for the generation capacity and the demand (that is, $\exists \bar x_{\max} \ge \bar x_i$, $\underline x_{\min} \le \underline x_i$, $d_{\min} \le d_i \le d_{\max}$ for all $i$), and (d) the desired precision $\epsilon$ is pre-determined.
Under these assumptions, the network operator can determine the worst cases of $b_{\boldsymbol{f}}$, $\sigma_2$, and $\delta$ in \eqref{eq:bark} as follows, so that $\bar k$ is computed and let $k=\bar k$ which is announced in public.
First, it follows that $b_{\boldsymbol{f}} \le \sqrt{N_{\max}} \max_{i,\lambda_i} |(dg_i/d\lambda)(\lambda_i)| \le \sqrt{N_{\max}} \max\{ |d_{\min} - \bar x_{\max}|, |d_{\max}-\underline x_{\min}|\}$.
Also, by \citep[Theorem 4.2]{M91}, any graph with unit weight on edges satisfying Assumption \ref{assmpt:graph} has the property that $\sigma_2 \ge 4/N_{\max}^2$.
Now, define 
\begin{equation*}
\frac{d\hat g_l}{d \lambda}(\lambda) = \begin{cases}
-\underline x_{\min}, & \lambda < \frac{d{\hat J}_l}{dx_l}(\underline x_{\min}), \\
-\hat v_l(\lambda), & \frac{d{\hat J}_l}{dx_l}(\underline x_{\min}) \le \lambda \le \frac{d{\hat J}_l}{dx_l}(\bar x_{\max}), \\
-\bar x_{\max}, & \frac{d{\hat J}_l}{dx_l}(\bar x_{\max}) < \lambda.
\end{cases}
\end{equation*}
for $l=1,\ldots,n$, where $\hat v_l$ is the inverse function of $(d{\hat J}_l/dx_l)$ on the corresponding interval.
Then, one can find $\hat \delta$, like in \eqref{eq:delta}, such that
\begin{equation*}
|a-b| \le \hat \delta \quad \Rightarrow \quad 
\left| \frac{d\hat g_l}{d\lambda}(a) - \frac{d\hat g_l}{d\lambda}(b) \right| \le \frac{\epsilon}{3N_{\max}}, 
\end{equation*}
for all $l = 1,\ldots,n$.
Since $J_i$ is one of $\hat J_l$, $l=1,\ldots,n$, it can be shown by the definitions of $(dg_i/d\lambda)$ and $(d\hat g_l/d\lambda)$ that 
$$\left| \frac{dg_i}{d\lambda}(a) - \frac{dg_i}{d\lambda}(b) \right| \le 
\max_{l=1,\ldots,n} \left| \frac{d\hat g_l}{d\lambda}(a) - \frac{d\hat g_l}{d\lambda}(b) \right|.$$
It then follows from \eqref{eq:delta} that $\hat \delta \le \delta$.
From the discussions so far, one can take
\begin{equation*}
k=\bar k = \frac{\max\{ |d_{\min} - \bar x_{\max}|, |d_{\max}-\underline x_{\min}|\} N_{\max}^{5/2}}{2 \hat \delta} \ge \frac{2b_{\boldsymbol{f}}}{\sigma_2\delta}.
\end{equation*}

\section{Simulation: IEEE 118 bus system}\label{sec:example}

We consider the IEEE 118 bus system\footnote{For more details about $J_i$, $d_i$, $\bar x_i$, $\underline x_i$, and the graph $\cG$, refer to \url{http://motor.ece.iit.edu/data/JEAS_IEEE118.doc}.} which consists of 118 nodes, 54 generators, 91 loads, and 186 branches. 
The local objective functions having generators are given by $J_i(x_i) = a_i + b_i x_i + c_i x_i^2$ whose coefficients have their values as $a_i \in [6.78,74.33]$, $b_i \in [8.3391,37.6968]$, and $c_i \in [0.0024,0.0697]$. 
The power demand of each node satisfies $d_i \in [0,277]$ and the total demand $\sum_{i=1}^N d_i$ is $3733.07$(MW). 
We assume that two nodes $(i,j)$ connected by a branch can communicate with each other in both directions.
By repeated simulations, we select the coupling gain $k=200$ for the distributed algorithm \eqref{eq:consensus_rule}.
The initial conditions $\lambda_i(0)$ are set, in this simulation, as $\lambda_i(0) = c_i(\bar{x}_i+\underline{x}_i)+b_i \in [\underline \lambda,\bar \lambda]$ in view of the fact that $\theta_i(\lambda_i(0)) = (\bar{x}_i+\underline{x}_i)/2$ is at the center of the possible generation range. 
For those $64$ nodes that have no generators, $(dg_i/d\lambda)(\lambda_i)$ of \eqref{eq:consensus_rule} becomes $d_i$, and \eqref{eq:gicases} reduces to $\theta_i(\cdot) \equiv 0$ as discussed in Remark \ref{rem:nogen1}. 

\begin{figure}
	\begin{center}
		\subfigure[Power generation $\theta_i(\lambda_i(t))$ of 118 nodes]{\includegraphics[width=8cm]{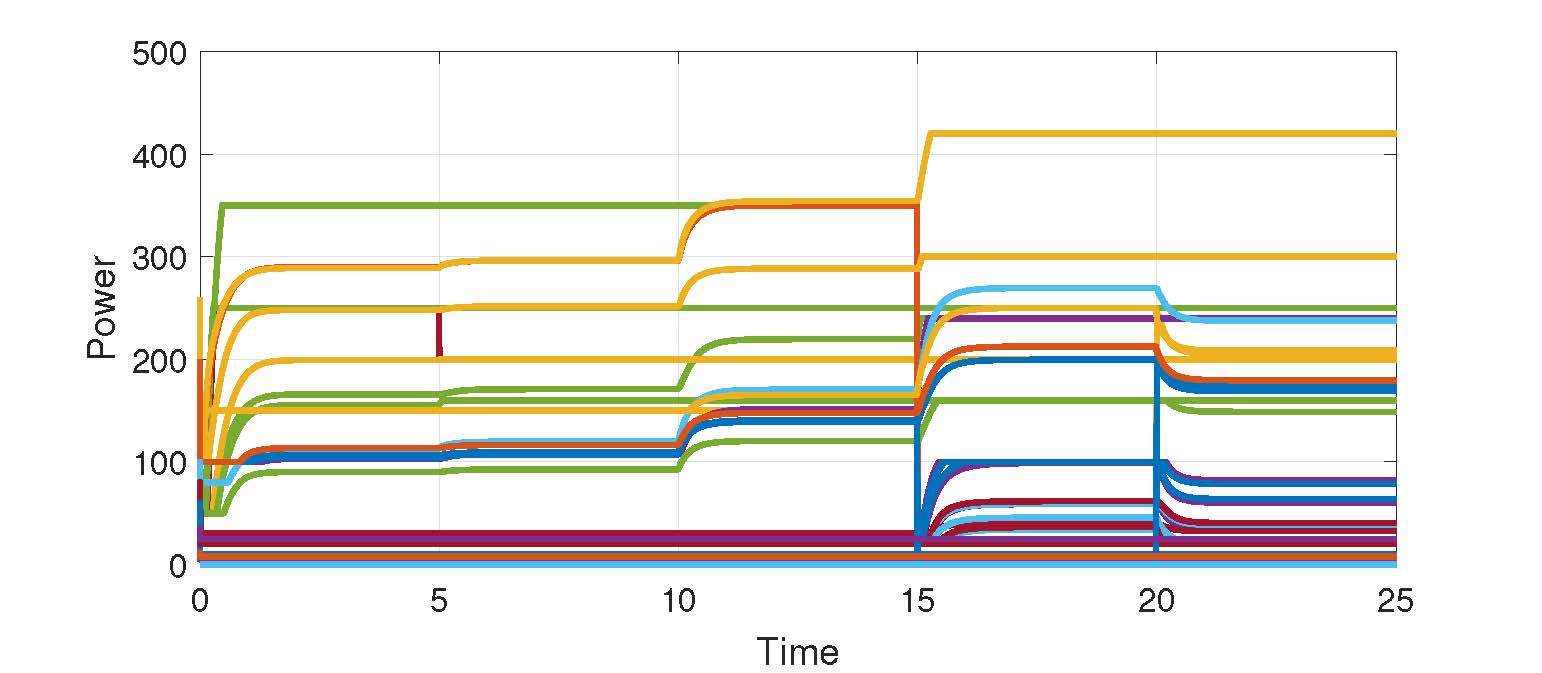}} \label{fig1_a}
		\subfigure[Optimal power generation $x_i^*(t)$ (for comparison)]{\includegraphics[width=8cm]
			{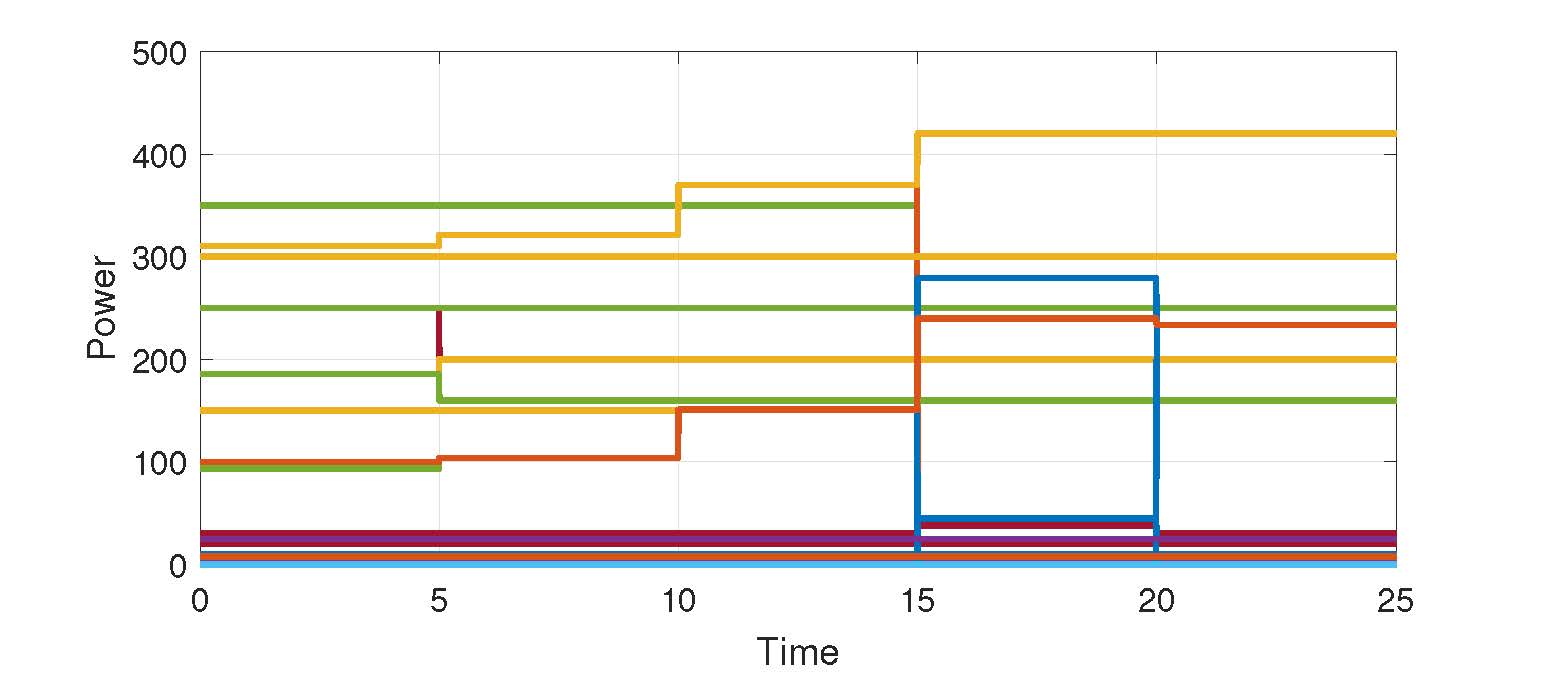}} \label{fig1_b}
		\subfigure[Power mismatch $x_i^*(t)-\theta_i(\lambda_i(t))$]{\includegraphics[width=8cm]
			{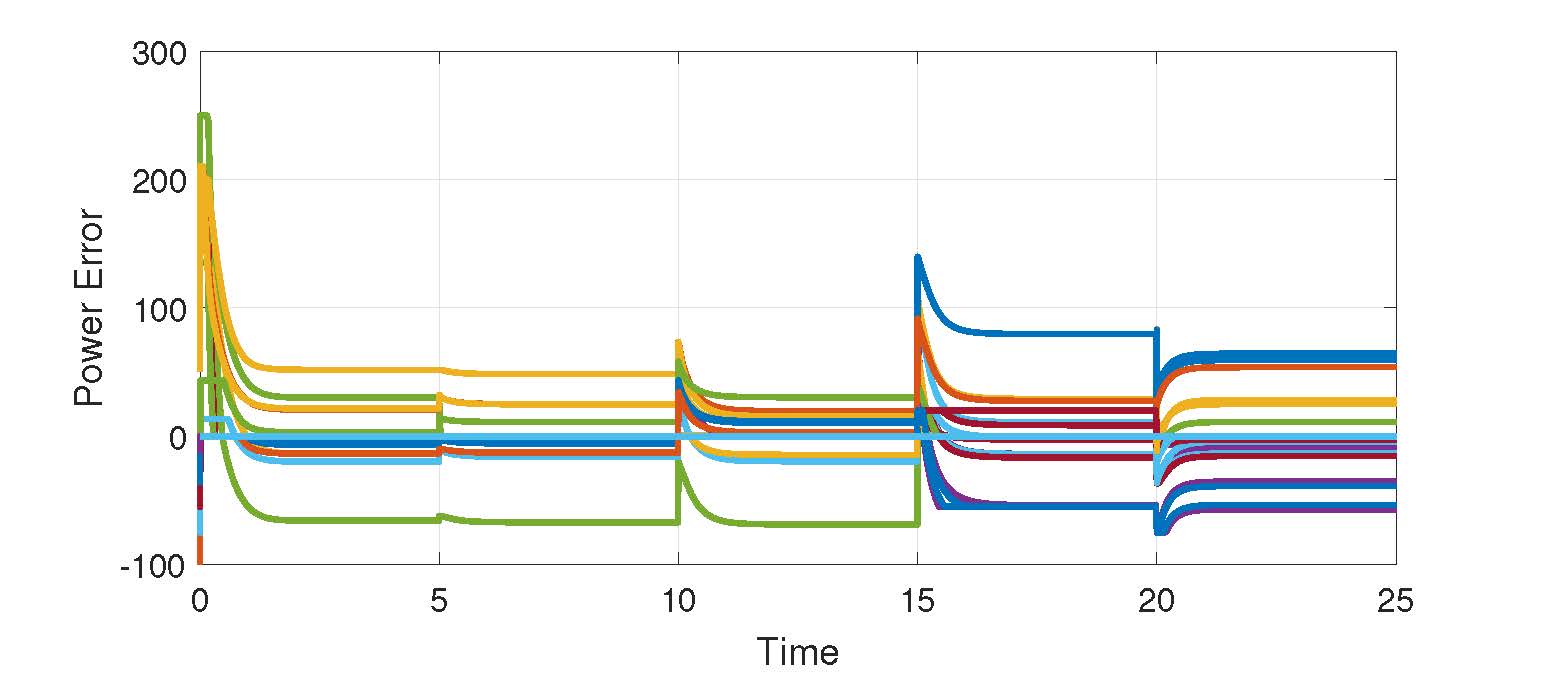}} \label{fig1_c}
		\subfigure[Total power mismatch $\sum_{i=1}^N (d_i - \theta_i(\lambda_i(t)))$]{\includegraphics[width=8cm]
			{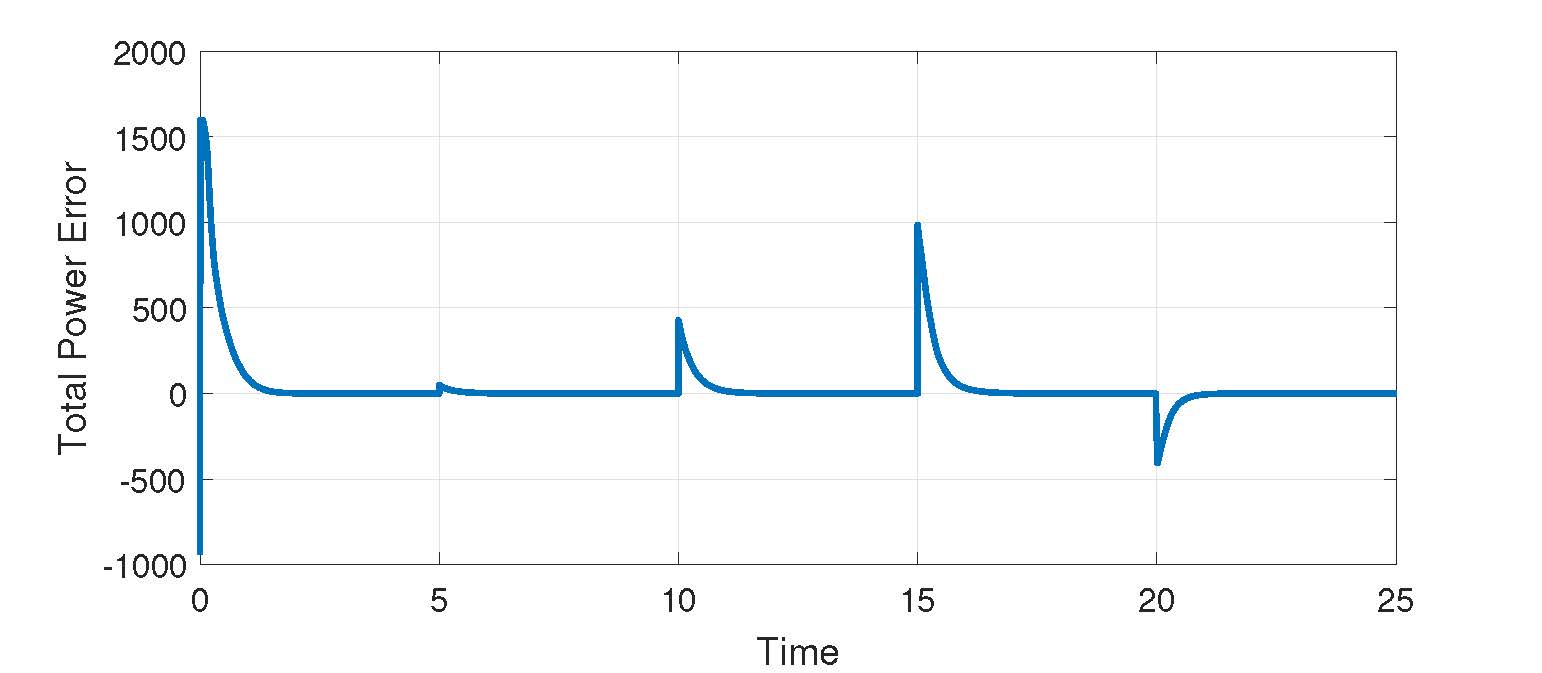}} \label{fig1_d}
		\subfigure[Total cost $\sum_{i=1}^N J_i(x_i^*(t))$ vs.~$\sum_{i=1}^N J_i(\theta_i(\lambda_i(t)))$]{\includegraphics[width=8cm]
			{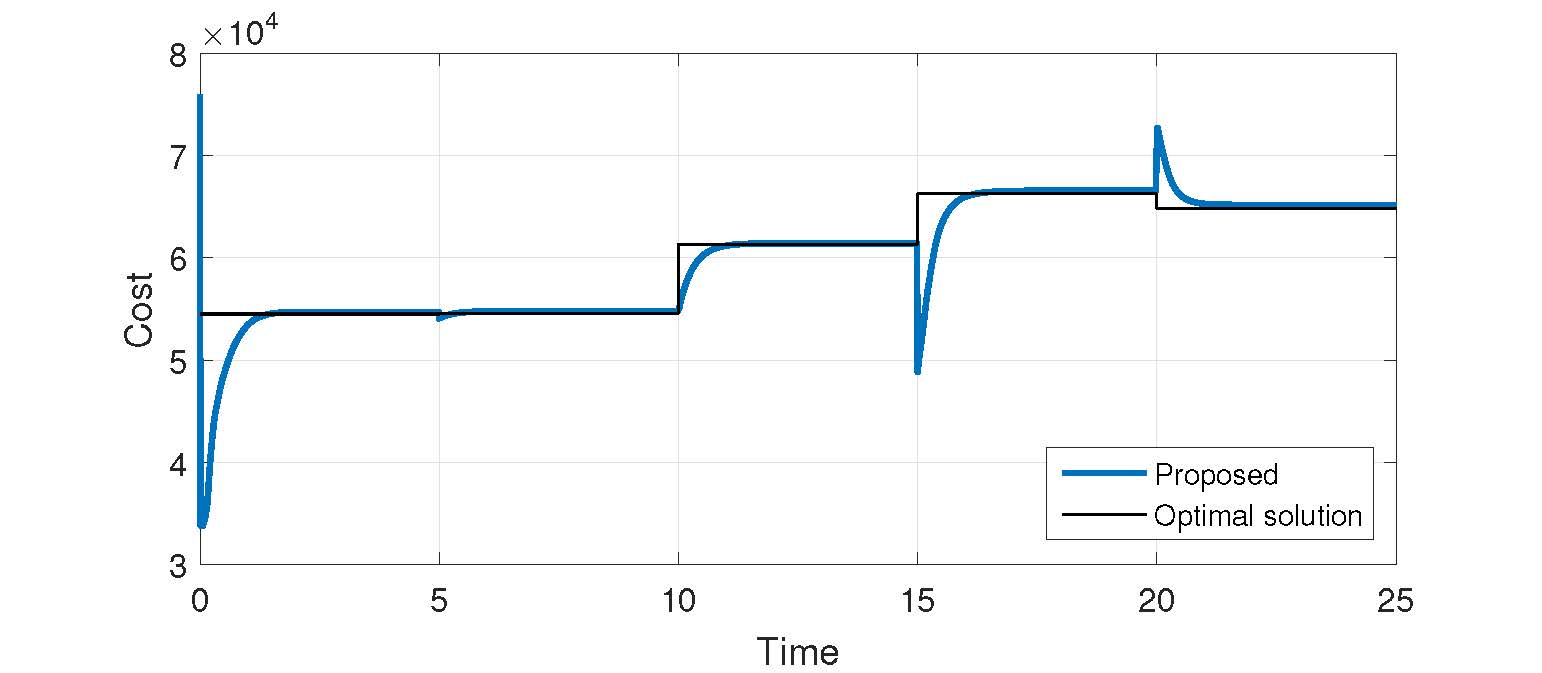}} \label{fig1_e}
	\end{center}
	\caption{Simulation result: Feasible case}\label{fig1}
\end{figure}

We consider the following scenarios to illustrate how the proposed algorithm works against the changes of DER, loads, and network topology:
\begin{enumerate}[(S1)]
	\item Change of DERs: At $t=5$s, ten generators change their upper limits of power generation by $-20\%$.
	\item Change of loads: At $t=10$s, ten nodes increase their loads (i.e., power demands) by $+40\%$ so that the total demand becomes $4161.41$(MW).
	\item Change of networks: At $t=15$s, nodes $10$, $26$, $65$, and $99$ stop generating power and the edges adjacent to them are removed. 
	We selected these four nodes since they have significant roles in power generation and/or network topology.
	\item Change of networks: At $t=20$s, nodes $10$ and $99$ restart generating power, and the edges adjacent to them are restored.
\end{enumerate}

Fig.~\ref{fig1} shows that the proposed algorithm can successfully obtain solutions of the EDP \eqref{eq:EDP} in a distributed manner. 
In particular, it can be seen from Fig.~\ref{fig1}(d) that the proposed algorithm maintains the power supply-demand balance even if the solution $\theta_i(\lambda_i(t))$ is sub-optimal approximation of $x_i^*(t)$ as seen in Fig.~\ref{fig1}(c).

Now, let us consider the following infeasibility cases to show that the proposed algorithm may allow to detect infeasibility in a distributed manner.
\begin{enumerate}[(S1)]
	\item Change of loads: At $t=5$s, node $1$ increases its power demand by $+4500$ so that the total demand becomes $8233.07$(MW).
	\item Change of loads: At $t=15$s, node $1$ decreases its power demand by $-4500$ so that the total demand recovers $3733.07$(MW).
\end{enumerate}

\begin{figure}
	\begin{center}
		\subfigure[$\lambda_i(t)$ in \eqref{eq:consensus_rule} of 118 nodes]{\includegraphics[width=8cm]{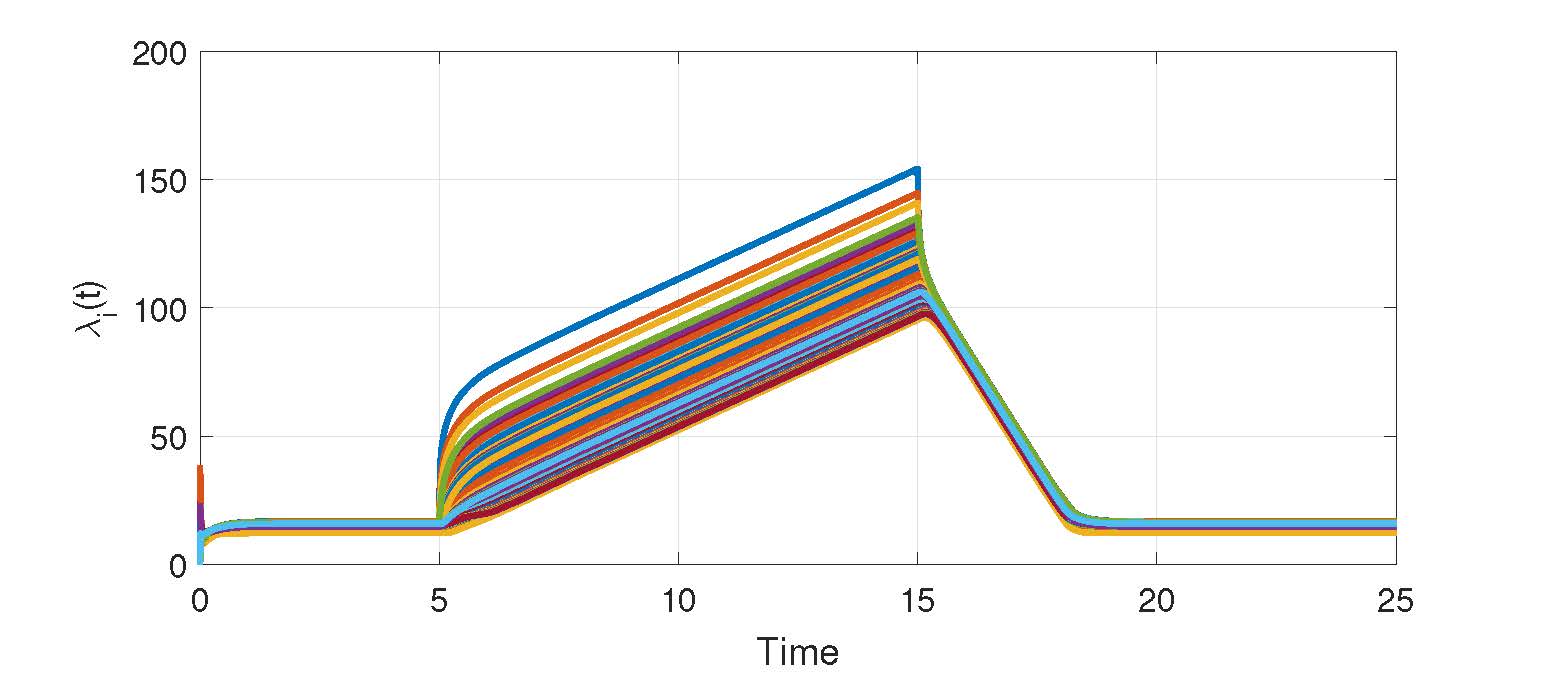}}
		\subfigure[$\dot\lambda_i(t)$ in \eqref{eq:consensus_rule} of 118 nodes]{\includegraphics[width=8cm]{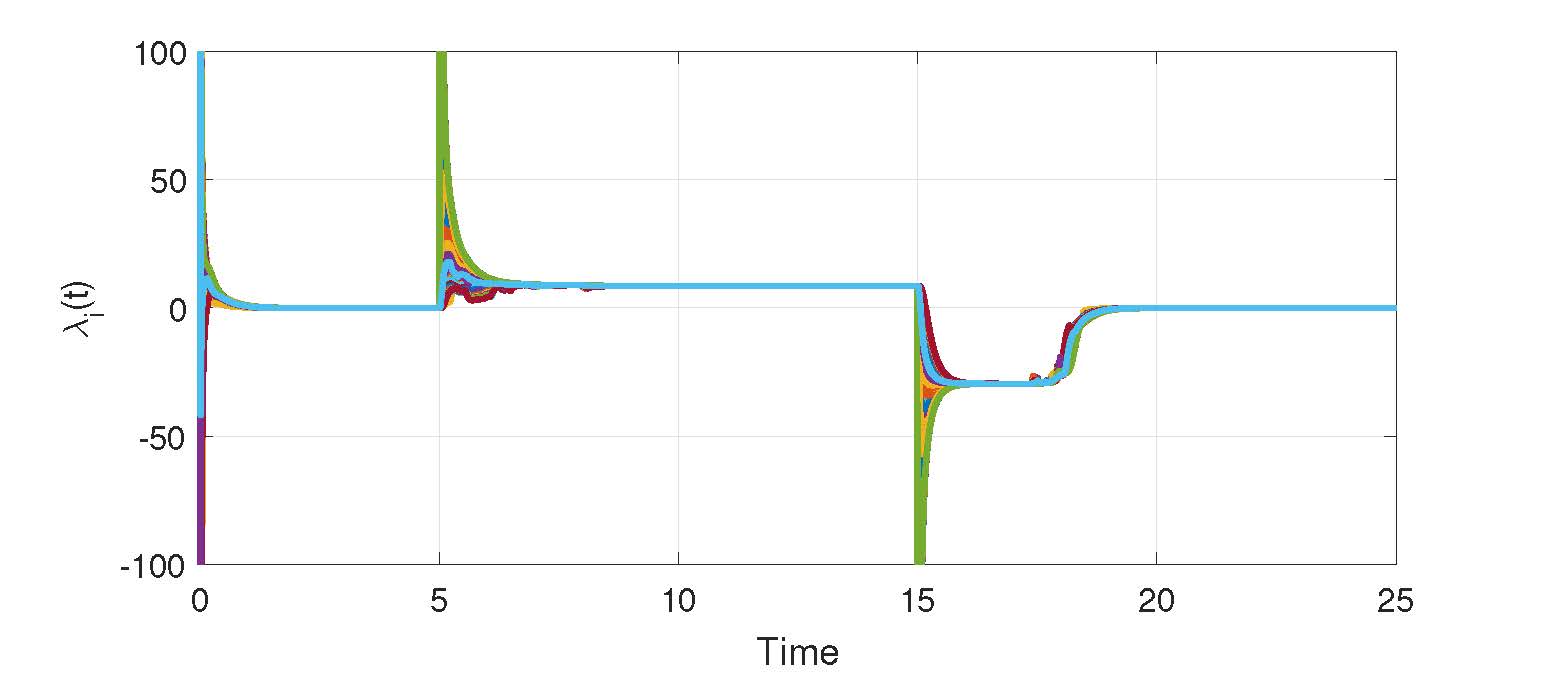}}
		\subfigure[$\dot\lambda_i(t)$, enlarged from (b) for $t \in \text{[4, 11]}$]{\includegraphics[width=8cm]{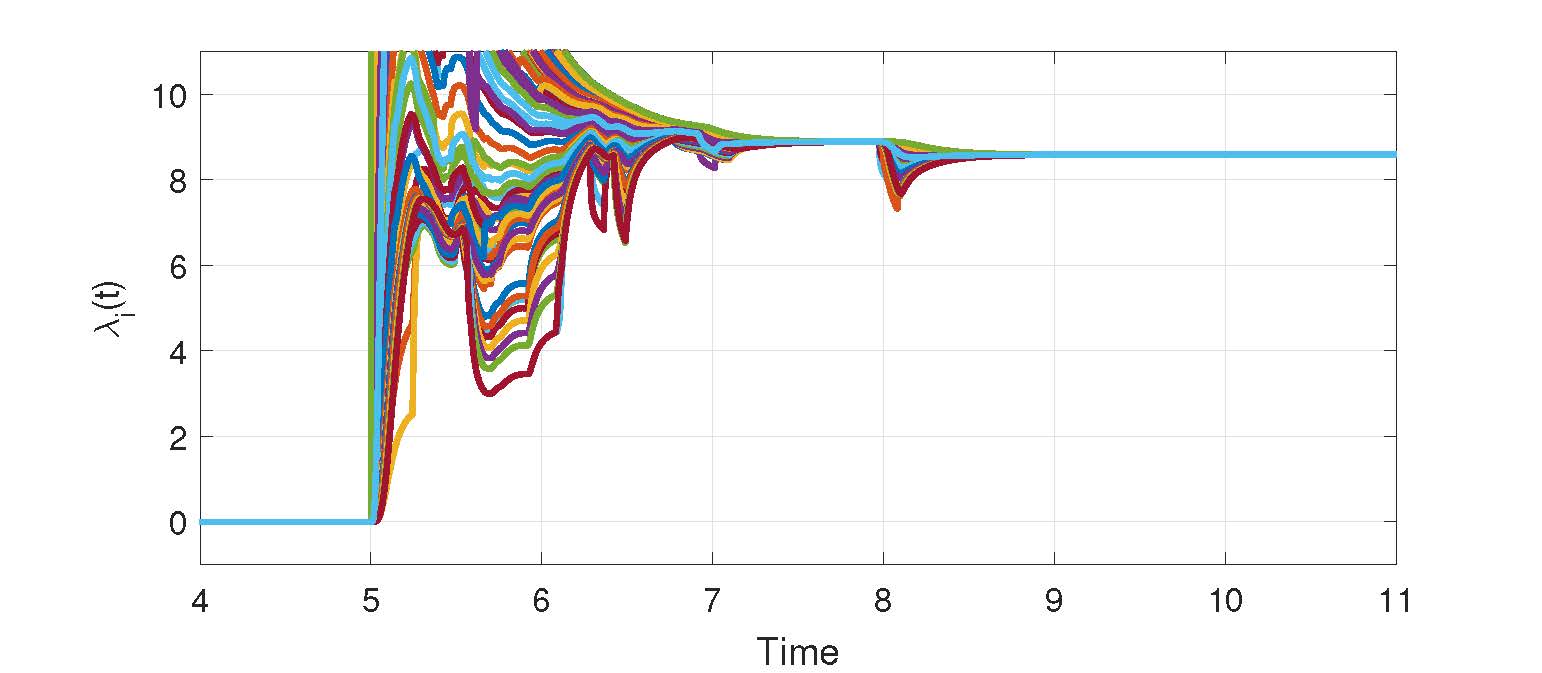}}
		\subfigure[Power generation $\theta_i(\lambda_i(t))$ of 118 nodes]{\includegraphics[width=8cm]{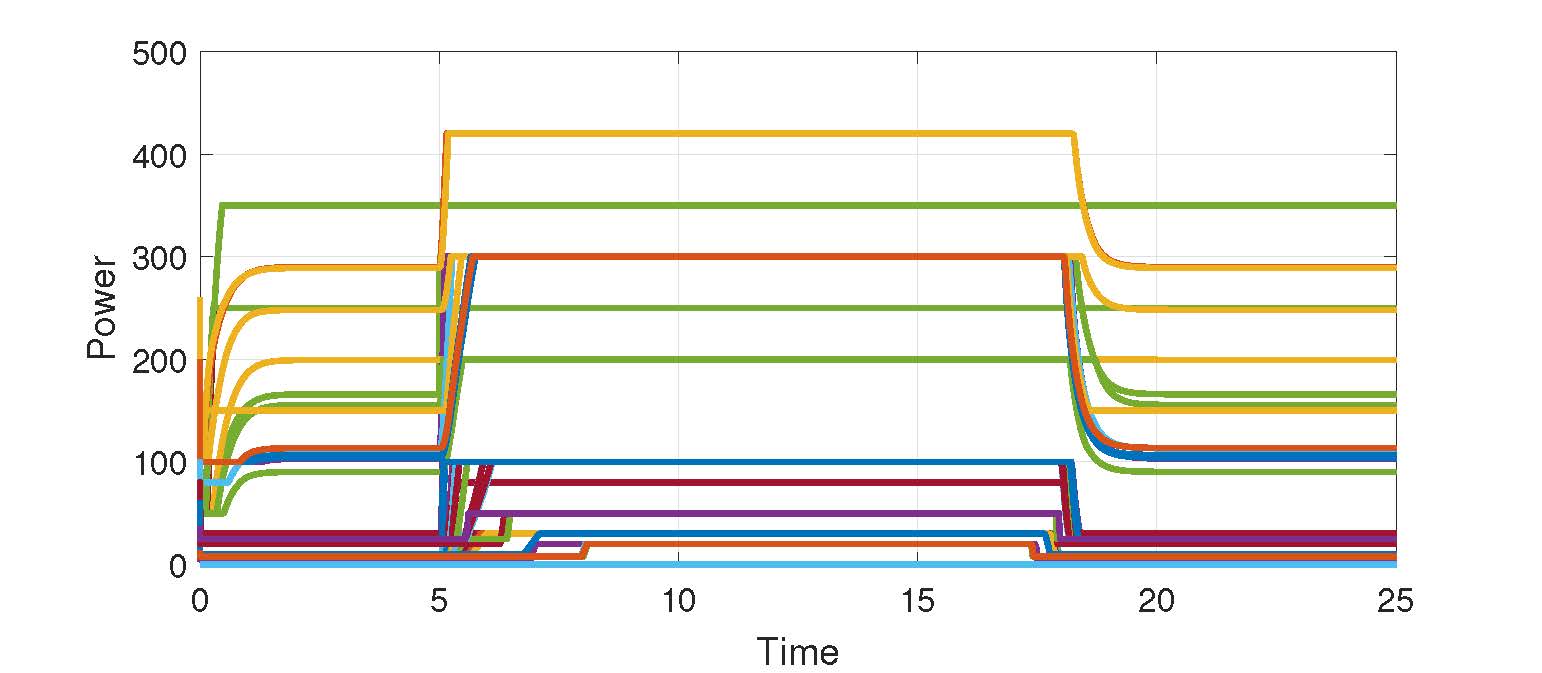}}
		\subfigure[Total power generation $\sum_{i=1}^N \theta_i(\lambda_i(t))$]{\includegraphics[width=8cm]{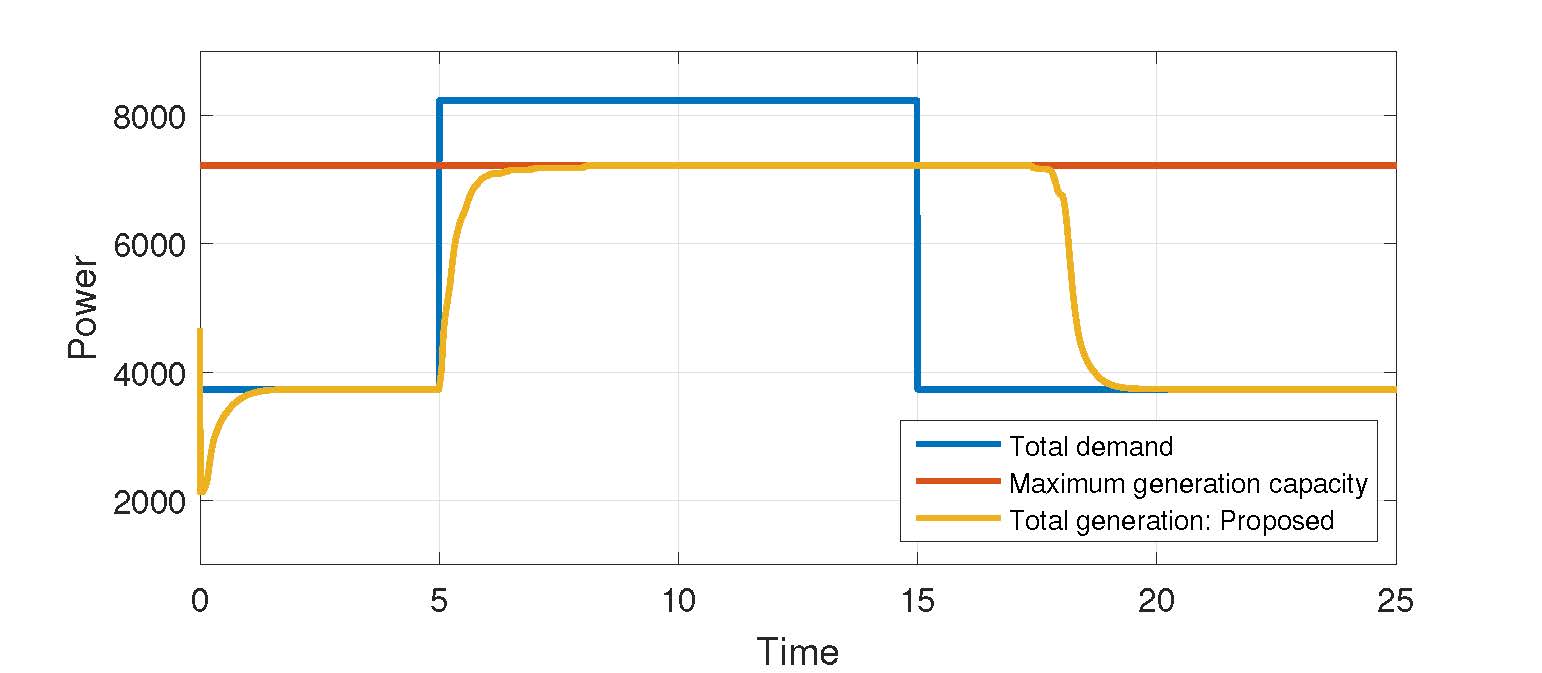}}
	\end{center}
	\caption{Simulation result: Infeasible case}\label{fig3}
\end{figure}

Fig.~\ref{fig3}(a) shows that $\lambda_i(t)$ tends to diverge to $\infty$ during $t \in [5,15]$ when the total demand $\sum_{i=1}^{N} d_i = 8233.07$ exceeds the maximum of power generation capacity $\sum_{i=1}^{N} \bar x_i = 7220$. 
In particular, it is noted from Fig.~\ref{fig3}(b)--(c) that $\dot \lambda_i(t)$ converges to the value $8.5853$ after $t=9$, which is exactly the value of $M_{\sf o} = (\sum_{i=1}^{N} d_i - \sum_{i=1}^{N} \bar x_i)/N = 8.5853$, as stated in Corollary \ref{thm:cs_inf}. 
Therefore, each node can figure out whether the infeasibility occurs and the amount of infeasibility.
Note from Fig.~\ref{fig3}(d)--(e) that all $\theta_i(\lambda_i(t))$ hit their maximum $\bar x_i$ during $t \in [5,15]$ which is within their generation capacities.
After the time $t=15$s, all nodes recover their feasible solutions even though the time to recover takes longer than in the normal operation.
Simulations are performed by the forward Euler discretization of \eqref{eq:consensus_rule} with the sampling period of 1ms.
	
\begin{ack}                               
The authors are grateful to anonymous reviewers for their motivating comments to consider the infeasible case and the guaranteed power balance, and to improve the presentation of the paper.
\end{ack}

\end{document}